\documentclass[fleqn,10pt]{wlscirep}
\usepackage[utf8]{inputenc}
\usepackage[T1]{fontenc}
\usepackage{lmodern}

\usepackage{subcaption}
\usepackage{setspace}

\usepackage{amsmath,amssymb,amsthm}
\usepackage{graphicx}
\usepackage{dcolumn}
\usepackage{bm}
\usepackage{hyperref}
\usepackage{siunitx}
\DeclareSIUnit\Molar{\textsc{M}}

\usepackage[normalem]{ulem}
\usepackage{comment}

\usepackage[capitalise]{cleveref}

\newcommand{\changed}[1]{#1}

\newcommand{\leish}{\textit{Leishmania\ }}
\newcommand{\lmex}{\textit{L.\ mexicana\ }}
\newcommand{\leishmex}{\textit{Leishmania mexicana\ }}
\newcommand{\leishns}{\textit{Leishmania}}
\newcommand{\lmexns}{\textit{L.\ mexicana}}
\newcommand{\leishmexns}{\textit{Leishmania mexicana}}
\newcommand{\tryp}{\textit{Trypanosomatidae\ }}
\newcommand{\trypns}{\textit{Trypanosomatidae}}

\title{Automated identification of flagella from videomicroscopy via the medial axis transform}

\author[1,*]{Benjamin J.\ Walker}
\author[1,2]{Kenta Ishimoto}
\author[3,4]{Richard J.\ Wheeler}
\affil[1]{Wolfson Centre for Mathematical Biology, Mathematical Institute, University of Oxford, Oxford, OX2 6GG, UK}
\affil[2]{Graduate School of Mathematical Sciences, The University of Tokyo, Tokyo, 153-8914, Japan}
\affil[3]{Nuffield Department of Medicine, University of Oxford, Oxford, OX3 7BN, UK}
\affil[4]{Sir William Dunn School of Pathology, University of Oxford, Oxford, OX1 3RE, UK}

\affil[*]{Benjamin.Walker@maths.ox.ac.uk}

\begin{abstract}

Ubiquitous in eukaryotic organisms, the flagellum is a well-studied organelle
that is well-known to be responsible for motility in a variety of organisms.
Commonly necessitated in their study is the capability to image and
subsequently track the movement of one or more flagella using videomicroscopy,
requiring digital isolation and location of the flagellum within a sequence of
frames. Such a process in general currently requires some researcher input,
providing some manual estimate or reliance on an experiment-specific heuristic
to correctly identify and track the motion of a flagellum. Here we present a
fully-automated method of flagellum identification from videomicroscopy based
on the fact that the flagella are of approximately constant width when viewed
by microscopy. We demonstrate the effectiveness of the algorithm by
application to captured videomicroscopy of \leishmexns{}, a parasitic
monoflagellate of the family \trypns{}. ImageJ Macros for flagellar
identification are provided, and high accuracy and remarkable throughput are
achieved via this unsupervised method, obtaining results comparable in quality
to previous studies of closely-related species but achieved without the need
for precursory measurements or the development of a specialised heuristic,
enabling in general the automated generation of digitised kinematic
descriptions of flagellar beating from videomicroscopy.

\end{abstract}

\begin{document}

\flushbottom
\maketitle
\thispagestyle{empty}

\section{Introduction}
\label{intro}
Well-studied and present across a wide range of organisms, the eukaryotic
flagellum is typically a long slender organelle that is known to perform a
variety of functional roles throughout nature, perhaps most notably in
spermatozoa where the beating of one or more flagella can render a gamete
motile \cite{Scharer2011}. Despite varying in function and also in
lengthscale, conserved across eukaryotic flagella is the characteristic `9+2'
axoneme \cite{Manton1952}. Formed of a central pair of singlet microtubules
surrounded by nine microtubule doublets, this cytoskeletal structure is
present along the length of the flagellum. This gives the organelle a
well-defined diameter, although this may be obscured or complicated by the
presence of accessory flagellar structures, such as the outer dense fibres of
the mammalian spermatozoon \cite{Fawcett1975}, or the paraflagellar rod
present in the parasitic family \tryp \cite{DeSouza1980}. Despite structural
additions to flagella, retained in typical optical videomicroscopy of
flagellated organisms is an approximately constant flagellar width, as is
exemplified in \Cref{fig:lmex}a.

Videomicroscopy of flagella at high resolution is pertinent to the study of
many microswimmers, such as the classically-investigated \textit{Crithidia
oncopelti} \cite{Holwill1974} and mammalian spermatozoon
\cite{Smith2009a,Ishijima2002,Ishijima2006,Ohmuro2006,Katz1978}, whether for
use in examining motility or exploring the mechanics of a beating flagellum.
In particular, obtaining a quantitative description of a flagellar beat
constitutes a key part of, and a barrier to, further research, with
significant human effort being dedicated to the problem. Researcher input
ranges from early manual approaches akin to those of Ishijima et al.\
\cite{Ishijima2002} and Vernon and Woolley \cite{Vernon2002,Vernon2004}, where
flagellum tracing is done by hand using acetate overlays, to semi-automated or
tailored methods \cite{Wan2014,Riedel-Kruse2007,Klindt2016,Mukundan2014}. The
work of Wan et al.\ \cite{Wan2014} and Klindt et al.\ \cite{Klindt2016} on the
alga \textit{Chlamydomonas} sees the flagellum extracted from
experimentally-fixed cells using custom software, with Smith et al.\
\cite{Smith2009a} utilising tuned image thresholding to ascertain the beating
pattern of a spermatozoan flagellum, exploiting the high contrast between the
spermatozoon body and flagellum specific to this microorganism and
experimental setup. Present in each of these methods is a significant need for
human input, be that in the calibration of a specialised algorithm or in the
manual processing of each captured image. Thus there is scope for a
fully-automated method for the identification and extraction of flagellar
kinematics from videomicroscopy, and any such scheme would be of relevance to
a wide scientific audience.

There are numerous semi and fully-automated methods for the digital tracking
of filaments, capable of accurately describing the kinematics of even
overlapping structures to sub-pixel precision, some examples being the
commonly-used `FIESTA' software \cite{Ruhnow2011}, the semi-automatic `Bohboh'
software (BohbohSoft, Tokyo, Japan)\cite{Baba1985}, the generalised linear
models of Xiao et al.\ \cite{Xiao2016a}, and the `active contour' methods of
Goldstein et al.\ \cite{Goldstein2010}, Hongsheng Li et al.\ \cite{Li2009}, Xu
et al.\ \cite{Xu2014}. Whilst invaluable in the tracking of free filaments,
such methods are less applicable to the study of a moving flagellated
microswimmer, where current methods may be unable to distinguish automatically
between the swimmer body and any attached flagella, with the exception of
specialised methods such as the probabilistic fitting approach of Yang et al.\
\cite{Yang2014} in application to spermatozoa. \Cref{fig:lmex}a exemplifies a
case where poor contrast between a flagellum and the rest of a cell may
prevent typical thresholding methods from distinguishing between the cell and
the flagellum, hence a different approach is needed to automatically
differentiate between them.

Image transforms, such as the Hough and Radon
transforms, are commonly used for feature extraction and image processing. A
pertinent example, the Hough transform is used primarily as a method for
detecting straight lines in binary images \cite{Duda1972,Hough1962}, and has
been applied to object tracking problems \cite{Nguyen2008}, whilst the related
Radon transform is widely used in tomographic reconstruction
\cite{Deans1983,Radon1986}. Less well-known than the Hough or Radon
transforms, the medial axis transform has seen use in studies of cellular
morphology \cite{Wheeler2012a,Blum1967}, with the transform encoding the approximate
size and shape of image features. Owing to the noted characteristic
shape of flagella, there is significant potential utility in this
shape-encoding image transform in the context of flagellum tracking and
identification.

Hence, in this paper we present a general fully-automatic method for the
identification of flagella in preprocessed videomicroscopy, focussing in
particular on the segmentation of an axoneme from a binary frame that includes
a free-swimming non-axonemal body. We exploit a feature of the axoneme biology
that is conserved across species and identifiable using the medial axis
transform, and apply the resulting automatic segmentation algorithm to a large
captured dataset of free-swimming \leishmex promastigotes, flagellated human
pathogens of the family \trypns. Evaluating our scheme against existing
semi-automatic software, we then demonstrate the applicability of the proposed
method to a spermatozoon dataset, showing multi-organism efficacy, and suggest
and comment on a number of possible refinements of the automated scheme,
considering also the performance of the scheme on reduced-quality datasets.

\section{Methods}
\label{methods}
\subsection{A common morphological trait}
\label{sec:feature}
As touched upon in \cref{intro}, the 9+2 cytoskeletal structure present across
eukaryotic flagella defines the slender cylindrical morphology of the
organelle \cite{Manton1952}. The presence of this underlying structure along
the entire length of the flagellum gives the filament a diameter that is
approximately constant at typical optical microscopy resolutions, even when
accessory structures are present. This is in stark contrast to what is in
general the more-varied visible morphology of the cell body, as can be seen in
\Cref{fig:lmex}, an example of a \lmex promastigote. Thus the flagellar
morphology may distinguish the slender organelle from the remainder of a cell,
and hence we will aim to isolate the flagellum from videomicroscopy by
identifying this feature, recognising it as a region of consistent visible
width.

\subsection{Medial axis transform}
\label{sec:medial}
\begin{figure}[ht]
\centering
  \includegraphics[width = 0.8\textwidth]{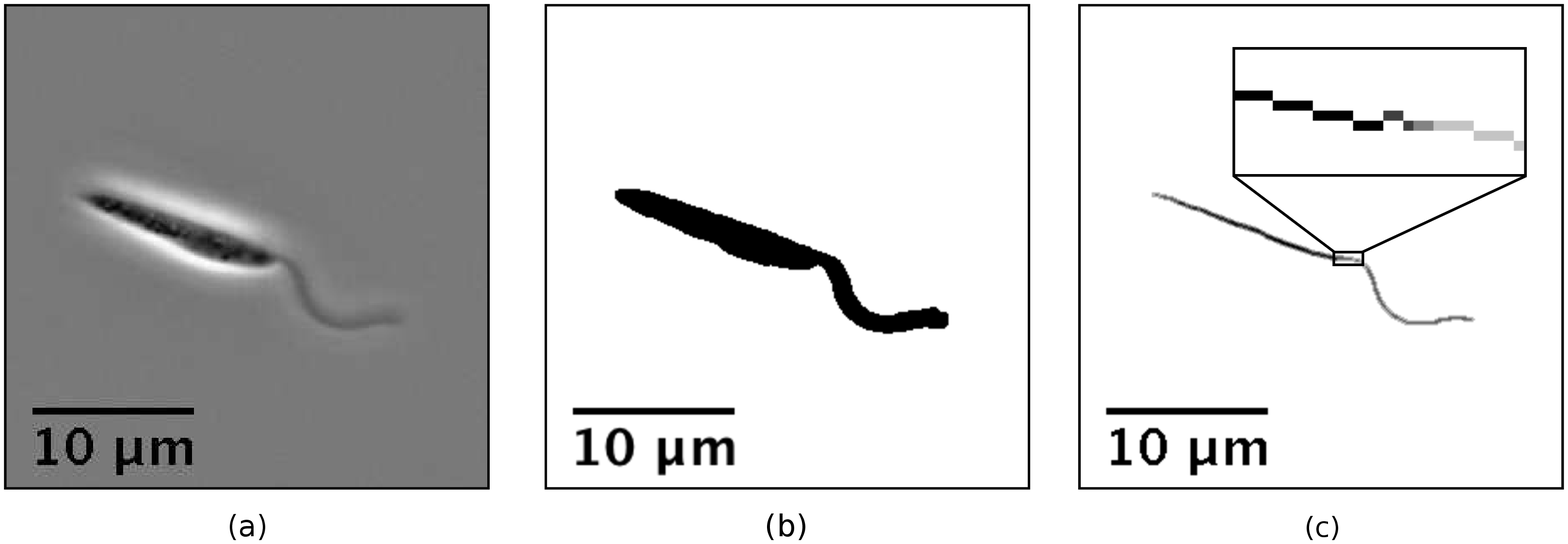}
	\caption{A sample frame taken from phase contrast videomicroscopy of a \lmex
  promastigote, from the dataset of Walker et al.\ \cite{Walker2018}. (a)
  Original frame. Despite the presence of an accessory structure, the
  paraflagellar rod, at the recorded resolution the flagellum appears to be of
  approximately-constant width. (b) Result of processing (a) into a binary
  image, following background subtraction and noise reduction.
  Existing methods of flagellum extraction are unable to automatically
  identify the flagellum in this image, typically requiring user input at this
  stage. (c) Result of the medial axis transform applied to (b), encoding the
  width of the cell along the medial line. Shown inset is an enlarged section
  of the transform.}
	\label{fig:lmex}
\end{figure}

\begin{figure}[ht]
\centering
  \includegraphics[width = 0.8\textwidth]{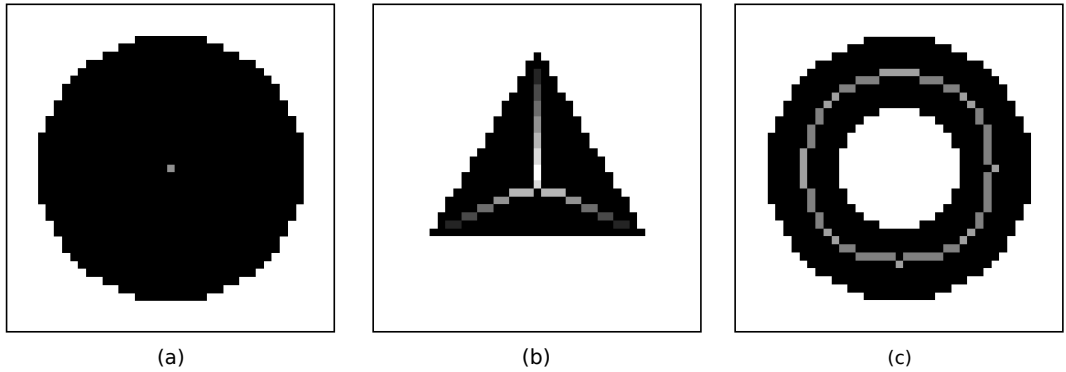}
  \caption{Examples of the medial axis transform applied to
  low-resolution simple shapes. The original image is shown in black, with the
  results of the transform superimposed in greyscale, where brighter pixels
  correspond to higher values of the transform and thus wider sections of the
  original shape. (a) A simple disk is mapped to a single point by the
  transform, with the radius of the disk encoded in the transform. (b) A
  triangular region is reduced to the skeleton shown, with the encoded width
  decreasing away from the skeleton and triangle's shared centre. (c) An
  annulus is transformed to a circle along its medial line, with a constant
  width-profile subject to artefacts of the rasterisation.}
  \label{fig:medial-ex}
\end{figure}

In order to locate regions of constant width and subsequently identify them as
axonemal filaments, we compute the medial axis transform of an image already
processed into a binary representation. The medial axis transform may be
thought of as the pixel-wise product of a Euclidean distance map and a
skeletonisation, with the distance map encoding at each point the shortest
distance to a point not included in the binary mask. As an example, consider
the binary image shown in \Cref{fig:lmex}b, obtained via a prefiltering and
image intensity thresholding of \Cref{fig:lmex}a, where the black region
identifies the cell body and flagellum. Skeletonising the region gives a curve
of the shape shown in \Cref{fig:lmex}c, where we are choosing to skeletonise
such that the skeleton of a connected region is also connected. Taking the
product of the resulting curve with a distance map of the original binary
image yields the grayscale image of \Cref{fig:lmex}c, with points of greater
magnitude (shown darker) corresponding to regions of greater width. Examples
of the medial axis transform applied to simple shapes are shown in
\Cref{fig:medial-ex}.

\subsection{Identifying the flagellar region}
\label{sec:ident}
\begin{figure}[ht]
\centering
  \includegraphics[width = 0.8\textwidth]{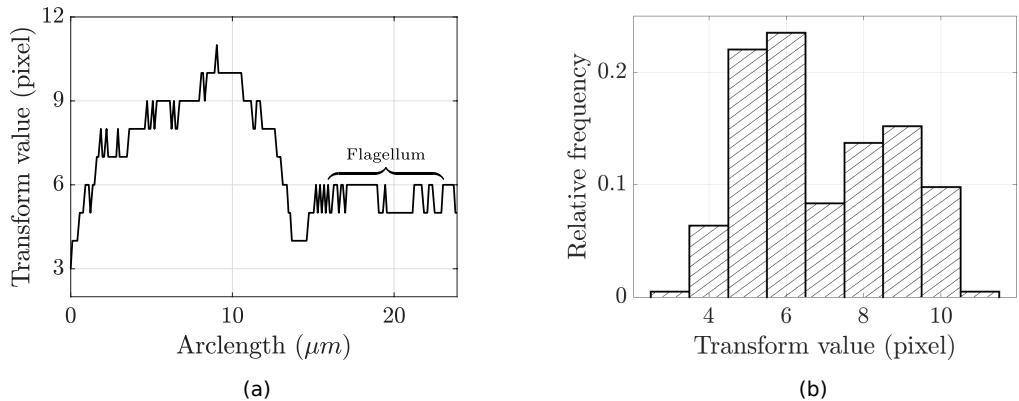}
	\caption{Analysis of a sample medial axis transform, corresponding to the
  cell in \Cref{fig:lmex}. (a) Values taken by the medial axis transform,
  shown in pixels, against the arclength, measured from left to right in
  \Cref{fig:lmex}. The flagellum may be clearly identified from this
  width-profile as the segment with approximately constant width, in contrast
  to the varied size of the rest of the cell. (b) A histogram of the discrete
  values of the transform shown in (a). A clear modal width can be seen around
  the flagellum width, suggesting that simple identification of the modal
  pixels may be sufficient to identify the flagellum in cases where the
  flagellar lengthscale is dominant.}
	\label{fig:medial}
\end{figure}

The results of the medial axis transform may be used to identify a region of
constant width, noted in \cref{sec:feature} to be a conserved morphological
feature of flagella. \Cref{fig:medial}a shows an example width-profile of a
flagellated swimmer, in particular the individual shown in \Cref{fig:lmex}.
With arclength being measured along the computed skeleton starting from the
left-most endpoint, a region of approximately constant width can be easily
identified, corresponding precisely to the flagellum. Analysis of the
derivatives of this profile with respect to arclength, for example examining
the variation in the first derivative, can be used to isolate the point of
flagellar attachment, and thus an endpoint of the flagellum, and subsequently
to segment the entire flagellum by the identification of this constant-width
region.

However, in many cases derivative analysis and sophisticated region
identification from the width-profile may be an unnecessary complication. As
indicated by the histogram of \Cref{fig:medial}b, we note that the observed
flagellar width found in \Cref{fig:medial}a corresponds to the approximate
mode of the width distribution of the cell. As the flagellum is typically a
long slender organelle, it may be reasoned to be a significant contributing
factor to a mode of the width distribution, and identified as such. Thus in
cases where flagellar length is relatively large it is sufficient to simply
identify those regions with cell width close to a modal value, subsequently
taking care to select the largest of such regions. In particular, such modal
analysis may be easily automated, and will be utilised in \Cref{calculation}.

\subsection{Procedure}
\label{sec:proc}
Hence we propose that the tracing of flagella from a single frame may be
achieved by the following overall procedure:
\begin{enumerate}[label=\roman*)]
	\item preprocess into a binary mask of the entire cell;
	\item perform and analyse a medial axis transform, isolating flagella by
	derivative or modal analysis;
	\item trace the resulting isolated filaments.
\end{enumerate}

Both the preprocessing and filament tracing steps may be achieved
automatically by use of existing tools, for example the methods of Ruhnow et
al.\ \cite{Ruhnow2011}, Xiao et al.\ \cite{Xiao2016a} and Goldstein et al.\
\cite{Goldstein2010}. The isolation of flagella via either the modal or
derivative analysis suggested in \cref{sec:ident} can also be perfomed without
user input, with an example of a simplistic but effective automated modal
analysis being provided in the Supplementary Information (Macro 1 and Macro
2). Choosing a particular method of flagellar isolation prior to image
processing, as we have done in Macro 1 and Macro 2, yields a fully-automatic
procedure for the identification of flagella from imaging data. Our presented
automatic implementations make use of a simple effective modal analysis of the
medial axis transform to accurately perform flagellar identification,
additionally locating the point of flagellar attachment. This choice of
implementation is motivated by \Cref{app:compare_modal_deriv}, where modal
analysis can be seen to outperform a simple derivative-based approach.

Efficient and effective application of the above method requires datasets of
suitable magnification, frame rate and contrast. In particular, magnification
must be sufficiently high to sharply capture the flagella, typically with
recorded flagellar width of at least 2/3 pixels. Frame rates should be high
enough to avoid motion blur, and contrast levels such that the organism may be
readily segmented from the background. \Cref{sec:low_quality_images} reports
in detail on the performance of Macro 1 when applied to lower-quality imaging
data, with an extension to datasets containing visually-overlapping flagella
being briefly considered in \Cref{app:overlap_pop}.

\subsection{Verification dataset}
\label{sec:verify}
In order to verify the proposed algorithm we will process the dataset
generated by Walker et al.\ \cite{Walker2018}. Comprising of approximately
150,000 frames combined from 126 \leishmex promastigotes, in general the
phase contrast videomicroscopy does not greatly differentiate the cell body
from the attached flagellum, with the flagellum not having a different grey
value to the cell body, thus existing thresholding approaches alone are not
sufficient for the digital isolation of the flagellum.

\subsection{Evaluation metrics}
\label{sec:metric}
To quantitatively assess the performance of any implementations against a
baseline result we will use the measures of missed detection rate (MDR) and
false detection rate (FDR) as used in the evaluation of the specialised
sperm-tracing scheme of Yang et al.\ \cite{Yang2014}. The MDR is the
percentage of pixels present in the baseline result that are more than $d$
pixels away from the segmented region produced by the test implementation, and
represents the proportion of the true region that has not been identified by
the test implementation. Conversely, the FDR is the percentage of pixels
identified by the test implementation that are more than $d$ pixels away from
the baseline result, and represents the prevalence of falsely-identified
regions. Here, following Yang et al.\ \cite{Yang2014}, we will take $d=3$
throughout unless otherwise stated\changed{, and note that an ideal
implementation would result in the MDR and the FDR both being zero}. We will
also report on the proportion of frames in a sample in which subjects have
been segmented successfully, where this is assessed qualitatively with
reference to by-eye identification.

\section{Results}
\label{calculation}
The computational procedure of \cref{sec:proc} was implemented in the ImageJ
macro language \cite{Schneider2012}. Motivated by
\Cref{app:compare_modal_deriv}, we opt in this instance to utilise the modal
analysis described in \cref{sec:ident}, creating a fully automated scheme that
is computationally-trivial but proves to be effective in practice. Sample
macros may be found in the Supplementary Information, where image
preprocessing and flagellum tracing have been implemented in a basic manner
that is sufficient for exemplifying the flagellar identification method.

\subsection{Flagellar identification in \lmexns}
\label{sec:ident_in_leish}
The 150,000 frames of the dataset of Walker et al.\ \cite{Walker2018} were
processed without user input using Macro 1 of the Supplementary Information, with
the overall computational time (including preprocessing and flagellum tracing)
being ca.\ 24 hours on a quad-core Intel\textsuperscript{\textregistered}
Core\texttrademark\ i7-6920HQ CPU running ImageJ 1.51u, with peak memory usage
no more than twice the size of a single frame.

To demonstrate the efficacy of the algorithm a 100-frame sample of the results
is presented in the Supplementary Information (Results 1), with the identified
flagellum centreline being highlighted in white on the original sample
(Dataset 1 of the Supplementary Information). In \Cref{fig:results}a we showcase
multiple frames of this composite as a montage, with the flagellum centreline
highlighted. Clear agreement can be seen between the segmented regions and
what may be identified by eye as flagella, obtaining similar results to what
may be expected of a manual tracing method but without any researcher input.
The same level of accuracy is present in an overwhelming majority of analysed
frames, with any loss only occuring due to the simplistic implementation of
flagellum tracing used here. Thus we have validated both an implementation of
and the methodology behind our proposed scheme of flagellar identification, in
addition to enabling a quantitative study of \lmex flagellar kinematics. 

Computed using the proposed automated method and as published by Walker et al.
\cite{Walker2018}, population-level beating statistics are given in
\Cref{table:beat_params} alongside those pertaining to the 100-frame \leish
sample. Reported in detail by Walker et al., from the analysis of this dataset
we may identify a characteristic sinusoidal form of the \leish flagellar beat.
\changed{In particular, this sinusoidal form can be clearly seen in
\Cref{fig:results}b, where captured flagella from multiple frames are shown
superimposed and from which a simple travelling wave form may be identified.
Indeed, as reported by Walker et al.\ \cite{Walker2018} Fourier analysis
identifies this periodic beating as having only a single prominent temporal
frequency, and thus a simple idealised functional form that enables
data-driven kinematic studies of the swimming behaviours of \leishns.}

\subsection{Evaluation against an existing semi-automatic method}
\label{sec:eval_against_bohboh}
We evaluate the pixel-wise performance of our implemented mode-based method
against the software package Bohboh (BohbohSoft, Tokyo, Japan)\cite{Baba1985},
which uses an established semi-automatic method of flagellum segmentation
\cite{Wood2005,Ishijima2002b,Ishijima2011,Miyata2015,Shiba2008a}. The
100-frame \leish sample as used in \Cref{sec:ident_in_leish} was processed
using Bohboh, requiring approximately 15 minutes of continued researcher input
and supervision, following which a TIFF containing the segmented flagellum
data was reconstructed from the output. Processing of the same sample was also
performed using Macro 1 of the Supplementary Information, which proceeded
automatically without any researcher input and completed within 15 seconds
using the computational resource described in \Cref{sec:ident_in_leish}.

Treating the reconstructed output of Bohboh as a baseline we compute the MDR
and FDR as defined in \Cref{sec:metric}, giving an MDR of 0.00\% and an FDR of
1.92\%. This minimal MDR corresponds to our implementation successfully
identifying all of the regions found in the baseline, and hence correctly
tracing the flagellum in each frame. The low FDR, comparable in magnitude to
the 1.78\% reported in the evaluation of the method of Yang et al.\
\cite{Yang2014}, approximately represents the inclusion of 1--3 additional
pixels per frame that are not present in the output of Bohboh, and thus
further demonstrates significant agreement between the output of our
implementation and that of Bohboh. Hence the results of our proposed
fully-automated methodology and implementation are in strong agreement with
the output of an established scheme. Further, our proposed method is
approximately 60 times faster than using the semi-automatic software, and
notably does not require any per-frame user input. Thus the proposed scheme is
seen to give results comparable to existing methods, but with greatly-reduced
user input and accompanying overall processing time.

\subsection{Application to a canonical flagellate}
To highlight the applicability of the method and implementation to a variety
of flagellated microorganisms we analyse a dataset of a swimming human
spermatozoon, specifically Supplementary Movie 1 of Ishimoto et al.\
\cite{Ishimoto2017b}. Example composite frames are shown in
\Cref{fig:results}c, and as in the case of \lmex demonstrate remarkable
accuracy, but we particularly note the presence of accessory structures to the
flagellum. We see that the midpiece of the flagellum has consistently not been
identified, owing to the greatly-increased width in this section, whilst the
principal piece is accurately segmented. However, the midpiece is itself of
approximately-constant visible width, thus we hypothesise that repeated modal
analysis will be able to correctly identify and even distinguish between these
different sections of the flagellum, and easily adapted from the current
implementation.

Processing was performed automatically using Macro 2 of the Supplementary
Material, which included slight adjustments to the preprocessing of Macro 1.
In this example some sections of the flagellum briefly move out of the plane
of imaging, which due to the simple preprocessing and filament tracing steps
implemented here results in some errors in flagellum identification, in
addition to variations in the final traced flagellum length. These are easily
corrected by the use of more-sophisticated approaches to these stages of
computation, such as those that have been previously commented upon in
\cref{sec:proc}. As an additional example, reconstruction of a non-planar
beating pattern may be performed as in Bukatin et al.\ \cite{Bukatin2015},
where image intensity is used to infer the 3-dimensional location of swimming
spermatozoa. Our approach may then be readily extended to 3 spatial
dimensions.

\subsection{Performance on low-quality imaging data}
\label{sec:low_quality_images}

In order to investigate the applicability of our methodology to a
range of imaging data we evaluate the effectiveness of our implementation on
intentionally-degraded datasets, emulating typical features of reduced-quality
imaging. To duplicates of the 100-frame \leish sample introduced in
\Cref{sec:ident_in_leish,sec:eval_against_bohboh} we perform a single image
degradation, with examples shown in \Cref{fig:degraded_montage}, and following
the application of Macro 1 of the Supplementary Information to the dataset we
compute the MDR and FDR relative to the baseline results established in
\Cref{sec:eval_against_bohboh}.

The results of this testing are presented in \Cref{table:degraded_results} for
a range of methods of image degradation mimicking common limitations of
microscale imaging quality, where we have in turn applied downsampling,
Gaussian noise, Gaussian blur, and gradient blending to the original sample
image. When downsampling we reduce the resolution in each spatial dimension by
a factor of 2 or 4, with successful segmentation still possible using Macro 1
in the case of a 2x reduction, whilst downsampling by a factor of 4 prevents
reliable segmentation.  Simulating higher framerate imaging, the effects of
applied Gaussian noise of standard deviation 10 (relative to pixel values in
the range 0-255) are minimal, with some loss of fidelity occurring near the
distal flagellar tip. Applying a Gaussian blur of radius 2 to emulate subject
defocussing or reduced resolution has little effect on the success of the
segmentation, with Macro 1 performing well relative to by-eye comparison.
Finally, blending the sample with a horizontal gradient to simulate
non-uniform illumination results in segmentation of good accuracy, with a loss
of precision around the flagellar attachment point that is expected to be
improved by additional preprocessing. Given this consistent acceptable
performance on all but the lowest quality data considered, the implementation
presented in Macro 1 is evidenced to be robust to a variety of reasonable
degradations in image quality, and hence is likely applicable to a range of
imaging data without significant modification or additional preprocessing.

\section{Discussion}
\label{results_discussion}
In this work we have presented and verified a method of flagellar
identification from the videomicroscopy of free-swimming bodies.
The formulation of a general method optimised for all flagellated
organisms or imaging contexts is not possible due to the diversity of
flagellated life, hence we have presented a high-quality and robust method for
addressing the basic case - that of a body with a single flagellum - which can
be readily adapted to a broad range of organisms and contexts. We note the
presence of a structural backbone ubiquitous in eukaryotic flagella and cilia,
giving these axonemal filaments a well-defined cross-section and therefore a
width. Additionally, we acknowledge that the presence of accessory structures
to the axoneme may significantly alter the ultrastructure of the organelle,
and therefore the width observed in videomicroscopy. However, the effects of
some such structures have been considered, as for the cases of the
paraflagellar rod of
\leishmex and the outer dense fibres of particular spermatozoa, and change in
visible width at typical optical resolutions and magnifications is either not
significant in general or may be further exploited to identify subsections of
a flagellum. Hence we proposed a procedure for the automated identification of
flagella-like structures from videomicroscopy that exploited the morphological
feature of consistent observed width, utilising the medial axis transform for
quantification.

To extract the location of a flagellum from the results of the medial axis
transform we have proposed two simple schemes: derivative analysis of the
width-profile, where consideration of local extrema of the first and second
derivatives is expected to be able to identify the proximal end of a flagellum,
and a less-complex modal analysis of the width distribution. Having
implemented and verified the latter approach, remarking that it is the most
suitable of the two, we note the scope for future work to refine the precise
methodology used here by performing more-sophisticated statistical analysis,
however we emphasise that even the basic scheme shown in this work was
sufficient for adequate identification of flagella. 

Our approach is suitable for cases where an image can be binarised
effectively, and we provide functional examples of preprocessing for typical
samples of cells swimming in a background-free environment. However, any
background subtraction or noise reduction approach could be used prior to
using the algorithm described here. Successful identification of a flagellum
here additionally depends upon the organelle being consistently contained
within the focal plane, and may be less successful in processing datasets in
which large out-of-plane deviations are consistently present - in the case of
organism which uses an extensively 3-dimensional flagellum motion, a
2-dimensional video does not contain the necessary data for a complete
analysis. Our approach is also likely unsuitable for use when processing
numerous cilia/flagella that may not be easily distinguished, such as those of
ciliated epithelia, but is in principle readily applicable to
multi-flagellated microorganisms such as \textit{Chlamydomonas}. In
\Cref{app:overlap_pop} we examined the potential effectiveness of the proposed
methodology on datasets in which flagella are observed to cross one another,
along with suggesting a suitable methodological refinement and thus
demonstrating potential applicability to dense flagellate populations. This
difficulty could be overcome in practice by diluting suspensions where
possible.

Preprocessing and flagellum tracing were implemented here in a basic manner,
and we recognise extensive scope for the incorporation of more-refined methods
of filament segmentation, such as the `FIESTA' suite of tools and active
contour methods of Ruhnow et al.\ \cite{Ruhnow2011} and Xiao et al.\
\cite{Xiao2016a} respectively, in addition to using established thresholding
methods for the creation of an initial binary mask. However, despite the
simplistic implementations used here, the proposed procedure was shown to
provide desirable accuracy at little computational or human cost, enabling
rapid quantification of flagellar motion in a sizeable dataset. Indeed, whilst
the presented approach identifies flagella in individual frames, the achieved
throughput is sufficiently high for use in video analysis, and has been
demonstrated to be robust to reductions in image quality.

A potential refinement to the proposed scheme involves combining our
morphological analysis with segmentation based on signal magnitudes, valuable
in cases where thresholding-based segmentation may be informative but only
partially applicable. We also suggest an application to the tracking of
free-swimming microorganisms, where the analysis of the width-profile may
yield the location of flagellar attachment and thus enable the tracking of
swimmer motion.

In summary, our proposed method for the identification of axonemal filaments
has been demonstrated to be of high accuracy and able to be implemented as a
fully-automated algorithm. We have verified our method on a large dataset
captured of a free-swimming flagellate, and achieved a high throughput and
accuracy without optimisation or refinement of a basic implementation. Notably
reliant only on a conserved morphological feature of axonemal filaments, the
considered procedure and implementations may potentially realise future study
of the kinematics of a wide range of flagellated and ciliated microorganisms.

\section*{Acknowledgements}
B.J.W.\ is supported by the UK Engineering and Physical Sciences Research
Council (EPSRC), grant EP/N509711/1. K.I.\ is supported by JSPS Overseas
Research Fellowship (29-0146), MEXT Leading Initiative for Excellent Young
Researchers (LEADER), and JSPS KAKENHI Grant Number JP18K13456. R.J.W.\ is
supported by the Wellcome Trust [103261/Z/13/Z,211075/Z/18/Z], with equipment
supported by a Wellcome Trust Investigator Award [104627/Z/14/Z].

\section*{Author contributions statement}
B.J.W.\ wrote the main manuscript text and prepared figures. K.I.\ performed
semi-automatic analysis on the 100-frame sample dataset. B.J.W., K.I.\ and
R.J.W.\ reviewed the manuscript.
\section*{Competing interests}
The authors declare no competing interests.

\section*{Data availability}
The full datasets generated during and/or analysed during the current study
are available from the corresponding author on reasonable request. Sample
codes and datasets are included in this published article (and its
Supplementary Information files).

\appendix
\section{Comparison of modal and derivative analysis}
\label{app:compare_modal_deriv}
We consider the viability of the derivative analysis suggested in
\Cref{sec:ident}, evaluating a simple implementation on the sample
\leish dataset provided in the Supplementary Information and comparing the
results against a modal analysis of the same width-profile. We will evaluate
performance on the simplest example of a flagellated swimmer, that with a
single flagellum that does not appear to intersect itself, with each method's
performance on this test case expected to be indicative of its performance in
more general settings. The modal analysis is implemented as in Macro 1 and
Macro 2 of the Supplementary Information, where the mode of the width-profile is
identified and then used as a simple threshold to isolate the image regions
corresponding to the flagellum.

Presented in Macro 3 of the Supplementary Information is a scheme based on a
simple derivative analysis, where the resulting skeleton of the medial axis
transform is traced from the flagellar tip and an approximation to the
derivative of the width-profile is computed via a forward difference operator.
At each point in the transform we determine if the approximated derivative is
an outlier in the context of the previously-computed derivatives, allowing for
noise due to the rasterised image and the discrete nature of the width-profile
and skeleton. Informed by the profile of \Cref{fig:medial}a, such an outlier
is assumed to correspond to the point of flagellar attachment, at which we
will have completely traced the portion of the skeleton corresponding to the
flagellum. This method of analysis is configurable via tuning of the foward
difference operator and the details of outlier identification, with a 3-step
forward difference operator being used here, found to give optimal
segmentation results for this test dataset.

On the 100-frame sample \leish dataset both methods of flagellum identification
achieve comparable pixel-wise accuracy to by-eye segmentation, with the
derivative-based method more-faithfully tracing the skeleton of the medial
axis transform than the modal thresholding method but the resulting difference
in output being less than 1\% of pixels, with typically only a single pixel
being affected. Though the derivative analysis gives a marginal increase in
pixel-wise accuracy over our implemented modal analysis, on the sample dataset
the derivative method correctly segments only 97\% of the given frames, in
comparison to the 100\% success rate of the modal analysis. Further, the
success of the derivative-based scheme is seen to be highly sensitive to the
details of the derivative approximation and method of outlier detection, with
the use of 2-step and 4-step foward difference operators correctly identifying
the approximate point of flagellar attachment in only 71\% and 95\% of frames
respectively. This sensitivity is expected to result in reduced general
applicability of the derivative-based approach in comparison to the modal
analysis, the latter having no comparable configurable parameters and noted to
perform satisfactory automated segmentation.

Thus we conclude that a simple modal analysis is better suited for application
to the test dataset than our derivative-based approach, with the robustness
and frame coverage of the former outweighing slight increases in pixel-wise
accuracy afforded by our derivative-based method. We therefore expect a modal
approach to be of the most utility in analysing large and diverse datasets,
where a robust and reliable method may be preferable to a highly-sensitive
scheme with marginally-improved accuracy.

\section{Adaptation to visually-overlapping flagella}
\label{app:overlap_pop}
Unseen in our analysed datasets but more prevalent in general, overlapping
flagella present possible complications to the application of our proposed
method of flagellum segmentation and tracing, requiring additional
consideration or more-sophisticated tracing methods. Here we briefly discuss
adaptations of the presented method to address two such complications, though
we emphasise that full consideration of complex flagellar interactions
requires significant future study.

\subsection{Effects on the observable width}
The apparent crossing of two flagella may intuitively be thought to increase
the observable width of any binary representation at the crossing point,
correspondingly increasing the value encoded in the medial axis transform
around that point. From manual examination of a range of artificial flagellar
intersections we remark that the crossings do not appear to significantly
increase the value of the transform in a large number of cases.

However, large variation may be observed when intersections are approximately
tangential, potentially doubling the observable width encoded in the
transform. Such cases may render the flagella indistinguishable from one
another, but in more-favourable cases will simply cause a large local increase
in the values of the transform. The modal implementation proposed in this
study would then typically result in disconnected components of the transform,
and thus would fail to identify the entire flagellar region.

We propose a simple augmentation to the existing methodology: the linear
stitching of any nearby isolated segments of the thresholded
transform to restore connectivity. Limiting the radius of reconnection to the
approximate flagellum width prevents the closure of any properly-segmented
regions, whilst enabling the joining of regions of flagellum overlap. Initial
evaluation of a simple implementation of this refinement step on artificial
test cases gives significantly improved accuracy, enabling segmentation of a range of crossing scenarios. However, this scheme is unable to
reliably segment approximately-tangential crossings, with future work needed to address this.

\subsection{Ill-defined transform arclength}
Crossing and self-intersection of flagella results in an ill-defined
transform arclength due to branching, potentially of issue to flagellum
segmentation and tracing. However, the considered and preferred modal analysis
is independent of transform topology and arclength, and thus only the
post-processing step of flagellum tracing is affected. In Macro 4 of the
Supplementary Information we provide a simple adaptation of the simple tracing
procedure used previously, resolving branches in the post-segmentation
skeleton by attempting to preserve the local flagellum tangent, appropriate for
all but the highest-curvature scenarios. We provide sample test cases in
Dataset 2 and Dataset 3 of two crossing skeletons, which are correctly traced
by Macro 4. Refinements of this sample approach may include
curvature-preserving branch resolution, though such work represents a large
field in image analysis research.


\begin{figure}[ht]
\centering
  \includegraphics[width = 0.8\textwidth]{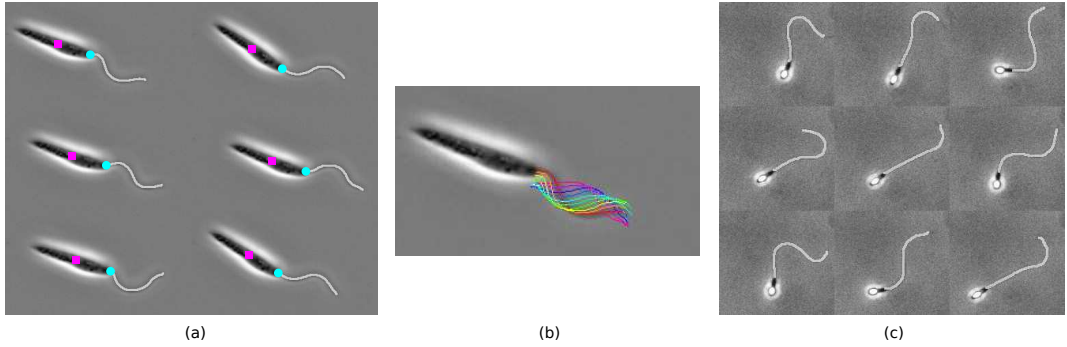}
	\caption{Composite data and results from an implementation of the proposed
  scheme. A montage of original frames with the identified flagella
  superimposed, showing very good agreement with identification by eye, for
  sample (a,b) \lmex and (c) spermatozoa. In (a) the computed cell centroids
  and locations of flagellar attachment are marked as magenta squares and cyan
  circles respectively. \changed{In (b) captured flagella from multiple frames
  are superimposed on the first frame of the sample dataset, with the simple
  sinusoidal nature of the flagellar beating being clearly visible.} Original
  frames from the datasets of Walker et al.\ \cite{Walker2018} (unpublished)
  and Ishimoto et al.\ \cite{Ishimoto2017b} respectively. Reprinted original
  frames of (b) with permission from [K. Ishimoto,
  H. Gad\^elha, E.A. Gaffney, D.J. Smith, J. Kirkman-Brown. Physical
  Review Letters 118, 124501, 2017] Copyright (2017) by the American
  Physical Society.}
	\label{fig:results}
\end{figure}

\begin{table}[ht]
  \centering
  \scriptsize
  \begin{tabular}{ccc}
    \hline
    Feature & Mean ($\pm$ S.D.) & Sample\\
    \hline
    Dominant frequency & $31.4\pm9.42$ \si{\hertz} & $29$ \si{\hertz}\\
    Flag.\ wavelength & $8.79\pm1.99$\si{\um} & $8.41$\si{\um}\\
    Flag.\ amplitude & $1.77\pm0.877$\si{\um} & $1.29$\si{\um}\\
    Flag.\ length & $14.9\pm4.90$\si{\um} & $10.6$\si{\um}\\
    \hline
  \end{tabular}
  \caption{Table of beat parameters extracted from the large \leish dataset of
  Walker et al.\ \cite{Walker2018}. Beating characteristics are computed
  following application of the proposed automated analysis, with averages
  shown for the entire population. To illustrate the results of analysis for a
  single cell, the beat parameters of the 100-frame \leish sample provided in
  the Supplementary Information are also presented. Population-level
  statistics reproduced from Walker et al.,
  doi:\href{https://doi.org/10.1016/j.jtbi.2018.11.016}{10.1016/j.jtbi.2018.11.016}
  under a Creative Commons Attribution License
  (\href{http://creativecommons.org/licenses/by/4.0/}{CC BY}), having been
  calculated via our proposed method of flagellar identification.
  \changed{With reference to the simple characteristic beating identified by
  Walker et al.\ using our proposed methodology, these features of the
  flagellar beating are sufficient to identify an idealised data-driven form
  of the \leish beat.}}
  \label{table:beat_params}
\end{table}

\begin{figure}[ht]
\centering
\includegraphics[width = 0.8\textwidth]{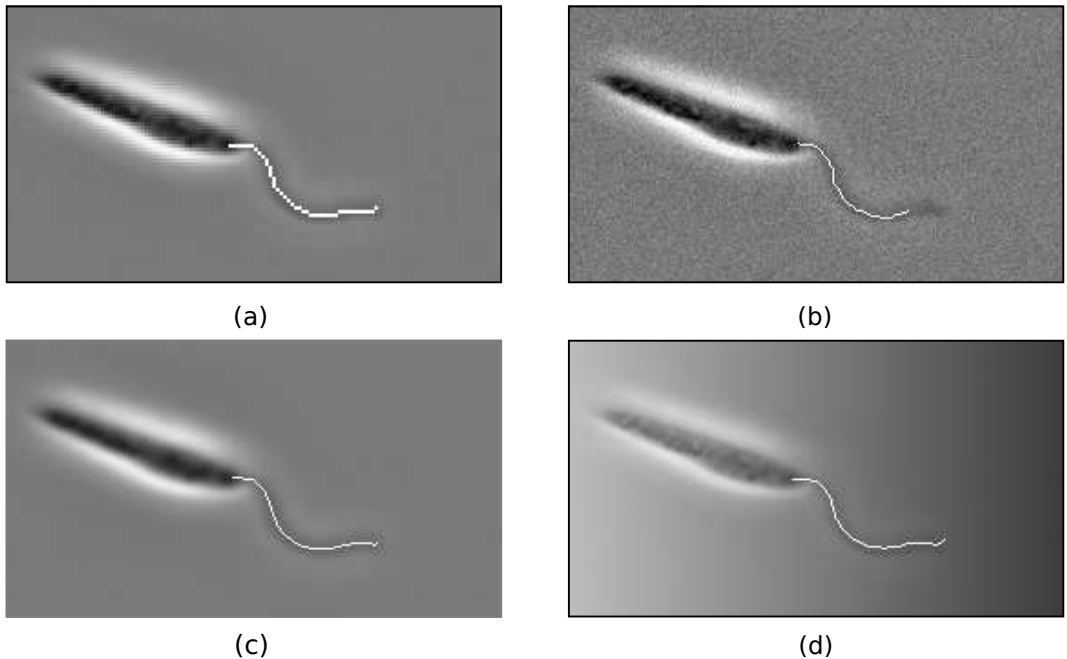}
\caption{Composite degraded data and results from an implementation of the
proposed scheme, where identified flagella are shown superimposed. Data has
been degraded by application of: (a) downsampling \changed{(2x)}; (b) Gaussian
noise; (c) Gaussian blur; (d) pixel-wise multiplication by a black-to-white
horizontal gradient. Acceptable flagellar segmentation by Macro 1 of the
Supplementary Material is seen for each degraded dataset shown here, with some
loss of accuracy at the tip or base of the flagellum. \changed{In particular,
Gaussian noise is seen in this case to prevent the distal tip of the flagellum
being segmented from the image background, a result of the basic preprocessing
implemented here and not characteristic in general of the proposed flagellum
segmentation procedure.}}
\label{fig:degraded_montage}
\end{figure}

\begin{table}[ht]{}
  \centering
  \scriptsize
  \begin{tabular}{cS[table-format=1.1]S[table-format=1.1]S[table-format=1.1]}
    \hline
    \multicolumn{1}{c}{Applied degradation} & \multicolumn{1}{c}{MDR (\%)} & \multicolumn{1}{c}{FDR (\%)} & \multicolumn{1}{c}{}{Success rate (\%)}\\
    \hline
    Downsampling (2x) & 0.14 & 2.21 & 100\\
    Downsampling (4x) & 14.47 & 47.48 & 33\\
    Gaussian noise & 1.45 & 1.57 & 99\\
    Gaussian blur & 0.01 & 2.21 & 100 \\
    Gradient blend & 0.02 & 6.12 & 100 \\
    \hline
  \end{tabular}
  \caption{Evaluated results of applying Macro 1 of the Supplementary Information
  to degraded imaging data. Downsampling by a factor of 2 in each spatial
  dimension, Gaussian blur of radius 2, and blending with a horizontal
  gradient each do not prohibit successful flagellum segmentation via Macro 1,
  with some loss of attachment point fidelity in the case of the gradient
  blend. Gaussian noise causes a loss of accuracy in the distal flagellar
  region, but satisfactory segmentation is retained. Downsampling by a factor
  of 4 prohibits successful segmentation via this non-specialised
  implementation, \changed{with the resulting image having dimensions of only
  68 x 38 pixels and hence being of very poor quality}. When computing the MDR
  and FDR pertaining to the downsampled images $d$ was taken to be 2 and 3 for
  the 2x and 4x downsampling respectively in order to account for the
  reduction in resolution.}
  \label{table:degraded_results}
\end{table}

\end{document}